\documentclass[12pt]{article}
\usepackage{graphicx}

\newcommand\pubnumber{DPF2013-32}
\newcommand\pubdate{\today}

\def\napoli{Fermi National Accelerator Laboratory\\
Batavia, Illinois, USA}
\def\support{\footnote{This work was supported by the Fermi National Accelerator Laboratory under US Department of Energy contract No. DE-AC02-07CH11359.}}

\def\Title#1{\begin{center} {\Large #1 } \end{center}}
\def\Author#1{\begin{center}{ \sc #1} \end{center}}
\def\Address#1{\begin{center}{ \it #1} \end{center}}

\newcommand\pubblock{\rightline{\begin{tabular}{l} \pubnumber\\
         \pubdate  \end{tabular}}}
\newenvironment{Abstract}{\begin{quotation}  }{\end{quotation}}
\newenvironment{Presented}{\begin{quotation} \begin{center} 
             PRESENTED AT\end{center}\bigskip 
      \begin{center}\begin{large}}{\end{large}\end{center} \end{quotation}}
\def\Acknowledgments{\bigskip  \bigskip \begin{center} \begin{large}
             \bf ACKNOWLEDGMENTS \end{large}\end{center}}

\def\beq{\begin{equation}}
\def\eeq#1{\label{#1}\end{equation}}
\def\eeqn{\end{equation}}

\def\beqa{\begin{eqnarray}}
\def\eeqa#1{\label{#1}\end{eqnarray}}
\def\eeqan{\end{eqnarray}}

\let\bar=\overbar

\def\Dslash{\not{\hbox{\kern-4pt $D$}}}
\def\dslash{\not{\hbox{\kern-2pt $\del$}}}

\def\msb{{\bar{\ssstyle M \kern -1pt S}}}

\begin{document}
\begin{titlepage}
\pubblock

\vfill
\Title{Neutrino Physics at DPF 2013}
\vfill
\Author{ Deborah A. Harris\support}
\Address{\napoli}
\vfill
\begin{Abstract}
The field of neutrino physics was covered at DPF 2013 in 32 talks, including three on theoretical advances and the remainder on experiments that spanned at least 17 different detectors.  This summary of those talks cannot do justice to the wealth of information presented, but will provide links to other material where possible.  There were allso two plenary session contributions on neutrino physics at this meeting: the current status of what we know about neutrino (oscillation) physics was outlined by Huber \cite{huberdpf}, and the next steps in long baseline oscillation expeirments were described by Fleming \cite{flemingdpf}.  This article covers a subset of the topics discussed at the meeting, with emphasis given to those talks that showed data or new results.  
\end{Abstract}
\vfill
\begin{Presented}
DPF 2013\\
The Meeting of the American Physical Society\\
Division of Particles and Fields\\
Santa Cruz, California, August 13--17, 2013\\
\end{Presented}
\vfill
\end{titlepage}
\def\thefootnote{\fnsymbol{footnote}}
\setcounter{footnote}{0}

\section{Introduction}

Although the field of neutrino oscillations currently gets a lot of attention, it is important to realize that a list of reasons to study neutrinos would extend far beyond the oscillation phenomenon.  One such list would include the following justifications:  

\begin{itemize} 
\item Neutrinos are among the most abundant particles in the universe
\item Neutrinos have been around since the universe was one second old
\item Neutrinos are signals from the highest energy accelerators in the universe
\item Neutrinos will tell us about how mass is generated
\item What other fermion could be its own antiparticle?
\item Neutrinos can see inside the nucleus like none slse
\item Neutrinos will tell us if we understand flavor
\item Neutrinos may be the reason that we enjoy such a healthy baryon asymmetry
\item Neutrinos broke the standard model and will tell us what is beyond
\end{itemize}

Finally, one last reason may include the fact that neutrinos can be studied across the widest energy ranges of all particles: from keV-level precision measurements of the neutrino mass, to EeV measurements of high energy cosmic neutrinos.  Talks inspired by most of the reasons listed above and covering the full range of energies were presented in the parallel session talks at DPF this year.  

\section{New Evidence for Astrophysical Neutrinos}

Whitehorn and Williams of the ICECUBE collaboration presented several interesting new results at DPF.  ICECUBE is a cubic kilometer of ice at the South pole that has been instrumented with 86 strings containing a total of 5160 phototubes.  The detector has the capability to distinguish between events that come from muon neutrino charged current interactions and other neutrino interactions, by looking at event topologies.  By looking for isolated large energy depositions without long tracks ICECUBE has seen the first evidence for electron neutrinos above 80~GeV.  The rates in that energy range are consistsent with those predicted to come from the atmosphere~\cite{whitehorn,williams}.  

More surprisingly, however, ICECUBE presented evidence at DPF 2013 that there are high energy neutrinos that are arriving at the detector at rates much higher than the rate expected from atmospheric neutrinos.  This implies that those high energy neutrinos are coming from astrophysical sources.  There was no evidence for clustering in the direction distribution of those high energy neutrinos but ICECUBE will continue to collect data for several more years to better understand these EeV neutrinos~\cite{whitehorn}.

\section{Absolute Neutrino Mass Measurements} 

Because neutrino masses are so much lower than the masses of any other fermions, this points to the possibility of a different mechanism of mass generation than interactions with the Higgs field.  For this reason alone it is critical to measure the absolute neutrino mass.  However, because neutrinos are so abundant in the universe the mass is also an important input into cosmological models of structure formation in the universe.  Although measurements of structure formation can give hints of the sum of the absolute neutrino masses, the ultimate goal is to see if the direct measurements and the cosmological measurements are in agreement.  

The current best limits on the absolute neutrino masses come from looking at the endpoint of the tritium beta decay spectrum.  The Mainz and Troitsk experiments have set limits of approximately 1.8~eV \cite{mainz,troitsk}.  The KATRIN experiment will extend this sensitivity by an order of magnitude, also by looking at the end point of the electron energy spectrum~\cite{KATRIN}.  

Oblath presented an update on Project 8, a new concept that could potentially measure the neutrino mass using Tritium decay but with superior projected precision above the current limits~\cite{oblath}.  The concept is that atomic tritium is trapped in a volume that is placed in a large uniform magnetic field.  When decay electrons are emitted, they will precess around the magnetic field lines, and emit photons with a characteristic cyclotron frequency.  With enough antennae around the volume and a quiet enough environment, that frequency can then be measured.  The frequency is a function of the electron's energy as follows:
\begin{equation}
\omega(\gamma) = \frac{eB}{K + m_e} 
\end{equation} 
where $eB$ is the electron's charge times the magnetic field,  $m_e$ and $K$ are the electron's mass and kinetic energy respectively.  Because of this relation the highest energy electrons will give the lowest frequencies, and they should be well separated from the lower energy electrons' signals.  RF technology is such that the expected frequency resolution is good enough for the forward- and backwards- Doppler shifted frequencies can even be distinguished from eachother, providing a clear signature.  

The first step in developing this concept is to prove that a magnetic field can be made uniform enough, and enough antennae can be placed in a quiet enough environment to extract a signal.  The Project 8 collaboration is currently working on proving these steps using $^{83m}Kr$.  This source will provide 18~KeV and 30~KeV conversion electrons as a signal.  In fact the electron should lose energy as it precesses around the magnetic field and gives off photons, and the prediction is that the analysis can see that energy loss as the electron precesses.  First krypton data were taken in January 2013 and analysis is underway.  

Kunde presented yet another new way to measure the neutrino mass.  The electron capture decay 
\begin{equation}
^{163}Ho + e^- \rightarrow ^{163}Dy^* + \nu_e 
\end{equation}
results in a small increase in energy of the Ho target, and that energy is measured with a transition edge sensor that goes from superconducting to normal conducting when the energy deposited is several eV~\cite{kunde}.  As with the Project 8 strategy, the first steps involve checks of the technique using a less rare element.  Using $^{55}Fe$ as a source of small energy depositions, this device demonstrated an energy resolution of $9.0\pm 0.2eV$ for that nucleus.  Some of that width is due to the atomic physics of the sample; it is expected that some improvements in resolution will come simply from using $^{163}Ho$ instead.  The plan is to do a test with  $^{163}Ho$  later this year~\cite{kunde}. 

\section{In search of a Fermion that is its own anti-particle}

A less direct but still important way to measure the neutrino mass is by looking at neutrino-less double beta decay.  This will not only provide a number which is again a linear superposition of neutrino masses, but will also, if seen, prove that neutrinos are Majorana particles, or their own antiparticles. 

There were two different techniques for a search for neutrino-less double beta decay presented at DPF 2013.  The EXO-200 experiment is using 200kg of liquid Xenon as a target, presented by Chaves and Kravitz~\cite{chaves,kravitz}, and the Majorana demonstrator is using 20kg of $^{78}Ge$, presented by Giovanetti~\cite{giovanetti}.  The two-neutrino double beta decay for the same isotope, while rare, still provides both a calibration source for the experiment and a potential background.  The expected EXO sensivitity is estimated to be 0.2eV, and that of the Majorana sensitivity is 0.1-0.2eV.  

At this meeting EXO-200 presented their measurement at the 2.85\% precision level of the two neutrino double beta decay lifetime for Xenon, which is the most precise measurement of this process to date~\cite{chaves}.  The electron energy spectrum shows good agreement of the shape with the known backgrounds, and it is clear that a new method to remove non-Xenon decay backgrounds is needed.  Progress has been made on a technique to tag the daughter barium in the decay, which would eliminate the non-Xenon backgrounds and improve the sensitivity of EXO-200 significantly below what was described above\cite{chaves,kravitz}. The MAJORANA demonstrator project is making steady progress at the Sanford underground laboratory.  In order to keep the target material free of cosmogenic activitation the time the target spends above ground is strictly controlled and reduced as much as possible~\cite{giovanetti}.  

\section{Using Neutrinos to probe the nucleus}

While the flavor sector of neutrinos is only recently being expanded, the use of neutrinos to probe the nucleus has a long history.  In previous years neutrino beams of ever higher energies were used to see inside the nucleus and probe ever shorter distances, all the way to determining valence quark distributions.  However, as oscilllation experiments use ever lower neutrino energy beams to study oscillation phenomena, these beams open up a new host of measurements, where rather than scattering off individual quarks in the nucleus, the scattering in a non-perturbative regime off entire nuclei become accessible at unprecedented statistics.  Nuclear physicicsts have joined neutrino experiments in hopes of seeing for example if the the nuclear environment has the same large effects on scattering phenomena as it has in charged lepton scattering.  

At the same time, physicists interested in measuring oscillation phoenomena are faced with the challenge of extracting precise oscillation parameters using neutrinos and anti-neutrinos scattering off of complex nuclei such as Carbon, Argon or Oxygen.  These phenomena must be measured as a function of neutrino energy, which can only be reconstructed event by event with the particles that leave the struck nucleus.  Also, even though we are in the era of large neutrino mixing, the oscillation probabilities are still at only at the several per cent level.  So even rare backgrounds must be well understood.  

Talks at this meeting showed progress towards a better understanding of neutrino interactions on nuclei, but it is also clear that this field has a long way to go before it reaches the precision needed for the ultimate oscillation measurements.  

The Argoneut experiment studies neutrino and antineutrino interactions on Argon using a TPC that was placed in the NuMI beamline just upstream of the MINOS Near Detector.  The fine granularity of the device allows for very low proton energy detection thresholds, and several event displays were shown which were consistent with two protons being emitted from a nucleon when a neutrino underwent a quasi-elastic interaction in the detector.  Szelc of Argoneut also showed a plot of proton multiplicity in their antineutrino data that was in slight disagreement with the prediction from the GIBUU neutrino event generator\cite{szelc}.  There is considerable work to be done to understand quantitatively the detector acceptance for these multi-proton events; Szelc and Grant presented two different plans for tests of Liquid Argon TPC's in a charged particle test beam \cite{szelc2,grant}.  

MINERvA studies neutrino and antineutrino interactions on a variety of different nuclei using a fine-grained solid scintillator active detector surrounded by electromagnetic and hadronic calorimetry.  As described by Rakotondravohitra, 
MINERvA has collected high statistics samples of neutrino and antineutrino quasi-elastic interactions and presented the cross sections as a function of momentum transferred to the nucleus ($Q^2$) as well as the distribution of energy near the vertices of these events~\cite{laza}.  Both of those observables show evidence that the simple model of nuclear effects, described by a relativistic Fermi gas of non-interacting particles, is not correct.  The MINERvA data are consistent with neutron-proton correlations in the nucleus: this would give additional two proton final states compared to a simple model, and additional two neutron final states in the case of the antineutrino mode~\cite{laza}.  

There is a long history of using charged lepton scattering measurements to understand the nucleus.  That history is now being more carefully utilized to make predictions for what neutrino scattering measurements might produce if the nuclear effects were similar for charged and neutral current exchanges.  In particular, charged lepton scattering measurements have seen an enhancement of the transversely polarized cross section.  Bodek presented an analysis of these data which make predictions on the quasi-elastic cross sections that neutrino experiments should see.  Those predictions are in better agreement with the world's data than other models which do not take into account this measured enhancement~\cite{bodek}.  

The T2K experiment is primarily designed to measure electron neutrino appearance over 295~km in a 700~MeV muon neutrino beam. To do that, however, the experiment uses a fine-grained near detector suite that also can probe the nucleus: specifically, the oxygen nucleus which is responsible for most of the events in the T2K far detector, the Super-Kamiokande Water Cerenkov detector.  The T2K near detector ND280 includes hydrocarbon targets (plastic scintillator) in addition to water targets, surrounded by TPC tracking and calorimetry and placed in a magnetic field provided by the UA2 magnet.  At this meeting there were progress reports by Adam on measurements of the small contamination of electron neutrinos in the muon neutrino beam~\cite{adam}, and by Hansen on measurements of the charged current quasi-elastic current cross section~\cite{hansen}.  

\section{Neutrino Flavor Physics}

Roughly half of the Neutrino parallel session talks at DPF2013 were on the topic of neutrino flavor measurements, so clearly this is an area of intense experimental activity.  Similar to the quark flavor sector, the neutrino flavor eigenstates are not the same as the mass eigenstates, and there is a (3$\times$3 unitary) mixing matrix that translates from one basis to the other.  Although there has been a long history of precision measurements to elucidate the quark mixing matrix, measurements in the neutrino sector are just starting.  Up until only 15 years ago the particle physics commmunity was still in denial about the possibility that neutrinos even had a non-zero mass, since this is not allowed without extensions to the Standard Model.  Through detailed measurements of solar, atmospheric, reactor, and most recently accelerator-produced neutrinos in the intervening years we now know the two independent mass splittings, and we know that two of the three mixing angles are relatively large, one of which is perhaps even 45 degrees which would mean maximal flavor mixing~\cite{huberdpf}.  Over this past year the field found definitive evidence that the last mixing angle to be measured, $\theta_{13}$, is also far from zero.  This latest discovery implies that measuring the CP-violating phase in the matrix is within the realm of the possible.  However, this latest discovery also implies that the field has a chance to verify that the framework itself is correct by looking for oscillation phenomena across many initial and final states, many energies and many baselines. 

\subsection{ In Search of the smallest mixing angle, $\theta_{13}$} 

The breadth of experimental conditions to observe the last mixing angle being non-zero was well-represented at the DPF2013 neutrino parallel session.  By looking at the disappearance over 1~km of MeV electron (anti-)neutrinos from reactors, Worcester and Carr reported on how liquid scintillator detectors can measure $\sin^22\theta_{13}$.  The Daya Bay experiment, using both near and far detectors, presented the most precise single measurement of the mixing angle itself~\cite{worcester}.  The Double Chooz experiment, which also uses liquid scintillator detectors near the Chooz nuclear reactor, is currently constructing its near detector but is also analyzing its far detector data and in particular can constrain the reactor off backgrounds because of the reactor complex's operations schedule~\cite{carr}.  

At the same time, accelerator-based experiments can also probe $\sin^22\theta_{13}$ by looking for electron neutrino appearance in 700MeV or 3.5GeV muon neutrino beams over 295~km or 735~km respectively.  Hignight of T2K presented updated results on 700~MeV electron neutrino appearance based on all the data taken through the summer of 2013~\cite{hignight}, and a new improved technique to eliminate backgrounds from neutral pions faking single electron events.  
The 28 events seen above the background of 4.6 exclude $\theta_{13}=0$ at 7.6 standard deviations.  Radovic of the MINOS experiment presented results for electron neutrino and antineutrino appearance at energies of 3GeV travelling over 735km~\cite{radovic}.  MINOS sees 152 signal events, 129 background in neutrino mode, and 20 signal events on 18.5 background events in antineutrino mode, in agreement with predictions based on reactor and T2K measurements.

In the future, the energy range where this phonemena is seen may be extended all the way to 10-100 GeV by looking at atmospheric neutrinos in the PINGU detector at the South Pole, as reported by Williams.  This proposal aims to be sensitive to $\theta_{13}$ and possibly even the mass hierarchy, given that other expeirments are showing that $\theta_{13}$ is several degrees~\cite{williams}.  

Although the GeV electron appearance measurements are consistent with the MeV reactor electron neutrino disappearance measurements, the field needs more precision in the appearance measurements to first of all test if the framework is correct and second to determine the mass hierarchy and search for CP violation.  T2K has only received 8\% of the planned exposure, and studies were presented by Friend comparing the ultimate physics reach of the T2K experiment for different choices of neutrino and antineutrino exposures over the total running period~\cite{friend}.  

The NOvA experiment, currently under construction in both Fermilab and in northern Minnesota, will test the framework by searching for electron neutrino appearance at a baseline of 810~km and with neutrinos with energies of about 2~GeV.  Sachdev presented studies done using a hit-level simulation of a new algorithm for predicting its electron neutrino backgrounds at the far detector~\cite{sachdev}, and Baird presented a study of the ultimate sensitivity to the largest mass splitting, by looking at muon neutrino disappearance~\cite{baird}.  Bian also presented the progress on the far detector construction~\cite{bian}.  
At the time of the DPF meeting NOvA had 18 out of the 28 blocks installed in the far detector hall, of which 11.49 were filled with liquid scintillator.  4.17 of the filled blocks were instrumented with electronics, and the near detector construction at Fermilab had also begun~\cite{bian}.

\subsection{Tests of the Oscillation Framework}

NOvA is one component of the worldwide program to determine if the framework for oscllations is even correct.  MicroBooNE, presented by Carls, is an experiment to look for evidence of a fourth (sterile) neutrino by searching for a mass squared difference that is consistent with that seen by LSND and MiniBooNE~\cite{collin}.  MicroBooNE will do this by searching for electron neutrino appearance in a 1~GeV muon neutrino beam traveling one kilometer.  MicroBooNE is currently under construction, and will elucidate not only the sterile neutrino sector, but will also pave the way to understanding how to scale up Liquid Argon detectors to the size that will be required for the next generation of neutrino experiments, specifically the Long Baseline Neutrino Experiment (LBNE).  

The most sensitive probes of the framework will come from comparisons across several neutrino energies.  LBNE will probe this by sending a broad band of neutrinos over 1300~km from Fermilab to South Dakota,   Shaevitz of the ISODAR collaboration presented two other techniques to probe the framework using low energy neutrinos from medical cyclotrons~\cite{shaevitz}.  

In fact a one test of the framework can be done by comparing all the evidence presented here on the largest mass squared difference.  While the MINOS experiment currently has the single most accurate measurement of that mass squared difference~\cite{radovic}, using neutrinos of about 3~GeV travelling over 735~km, other measurements of the mass squared splitting were presented from T2K using 700~MeV neutrinos, and even from ICECUBE, comparing the angular distributions of 20-100~GeV neutrinos to those of neutrinos of energies above 100~GeV~\cite{williams}.  
In future meetings we may be able to compare the mass squared splitting measurements from both 700~MeV, 2~GeV and 1~MeV neutrinos, once T2K~\cite{friend}, NOvA~\cite{baird}, and Daya Bay~\cite{worcester} achieve the currently planned statistics.  

\subsection{In Search of CP violation in the Lepton Sector} 

There is a large range of ideas going forward to determine the mass hieararchy and the CP-violating phase.  Although many experiments may only be sensitive enough to see at two sigma or so a "preference" for a specific mass hierarchy or CP violating phase, we should not forget what statistical tests really mean.  
We want to see these phenomena in as many ways as possible, and the field should not stop thinking about ways to improve on current plans.  One experiment was not enough in the case of CP violation in the quark sector, and the same is true for the lepton sector.  

LBNE is gaining momentum, collaborators, and sophistication in its analysis techniques. Guardincerri presented new advances in beamline instrumentation which are aimed at {\em in situ} measurements of the muon flux produced simultaneously with the neutrino flux~\cite{guardincerri}.    Studies of the ultimate LBNE sensivity were presented by Bass with new more sohpisticated treatments of detector capabilities, and the uncertainties associated both with neutrino fluxes and neutrino cross sections~\cite{bass}.

\section{In search of more surprises} 

In our enthusiasm of trying to figure out the best ways to get to the mass hierarchy and CP violation, we must not forget to make the most of the data sets that we do have in hand to look for other surprises in neutrino physics.  One need only consider the history of neutrinos to know that surprises come where they are least expected.  Towards that end Spitz of the Double Chooz experiment presented results from a search for Lorentz Violation by looking for a time dependence in the reactor neutrino rate.  Similarly, by looking at the kinematic distributions of the Double Chooz data, one can also search for neutron antineutron oscillations~\cite{spitz}.  Finally, Zhuridov presented constraints on the neutrino magnetic moment by considering energy loss in globular cluster red giants, and also from experimental results on neutrino electron scattering.  He also presented a "democratic neutrino" theory that agrees with SNO and Atmospheric neutrino results and makes predictions for low energy beta decays, magnetic moments, and neutrino-less double beta decay experiments~\cite{zhuridov}.  Ahmed presented constraints on the charged Higgs and W' sector by considering $\tau\rightarrow\pi\nu_\tau$ and  $\tau\rightarrow\rho\nu_\tau$ branching ratios.  The $\nu_\tau$ signal in the Super-Kamiokande atmospheric neutrino analysis was also used to set constraints on those sectors~\cite{ahmed}.  

Radovic of the MINOS experiment presented results of a 
search for sterile neutrinos by comparing neutral current events in both the near and far detectors.  A sterile neutrino would not have neutral current interactions, yet all the active neutrinos do have neutral current interactions, at equal rates.  This is similar to the SNO technique which compared neutral current processes to charged current processes to test solar neutrino oscillations~\cite{SNO}.  MINOS+, which is just starting to take data now, will have improved sensitivity to sterile neutrino sector through muon neutrino disappearance just above the oscillation minimum~\cite{radovic}.  

\section{Conclusions}

It is clear from the talks presented at DPF 2013 that this is an extremely exciting time to be studying the physics of neutrinos.  There are currently running experiments that are looking at energies from a few keV to a few EeV to address a very broad range of physics.  The progress to get to this point is particularly fast if one considers how recently the field was in denial of the possibility that neutrinos has non-zero mass in the first place.  In particular, there are several new ideas that were discussed at this meeting on how best to measure the absolute neutrino mass, which is something we will ultimately need if we are to understand where neutrinos get mass from in the first place.  There are several rich data sets being mined to understand precisely how neutrinos at GeV energies interact, which will tell us not only about the nucleus but enable the field ultimately to measure CP violation and the neutrino mass hierarchy.  Finally, there is an exciting complementarity and flexibility in the current program of long baseline oscillation experiments.  At this meeting we saw several examples of collaborations studying how to take advantage of that complementarity to get the farthest in the underlying physics. 

\Acknowledgments

I would like to thank all the speakers at the DPF 2013 neutrino parallel sessions, and the session organizers Matt Toups and Patrick Huber for providing such a broad snapshot of the field.


\begin{thebibliography}{99}

\bibitem{huberdpf}
  A.~de Gouvea {\it et al.}  [Intensity Frontier Neutrino Working Group Collaboration],
  arXiv:1310.4340 [hep-ex].
\bibitem{flemingdpf} B. Fleming, DPF Presentation ID 14
\bibitem{whitehorn}  N. Whitehorn, (ICECUBE) DPF2013 presentation ID 283
\bibitem{williams} D. Williams, DPF Presentation ID 269 and 
  M.~G.~Aartsen {\it et al.}  [IceCube and PINGU Collaborations],
  arXiv:1306.5846 [astro-ph.IM].
\bibitem{mainz} C. Kraus {\em et al.} Eur. Phys. J.
{\bf C40} 447–468 \verb+(hep-ex/0412056)+ 2005
\bibitem{troitsk} 
  V.~N.~Aseev {\it et al.}  [Troitsk Collaboration],
  Phys.\ Rev.\ D {\bf 84}, 112003 (2011)
  [arXiv:1108.5034 [hep-ex]].
\bibitem{KATRIN}     [for the KATRIN Collaboration],
  arXiv:1307.5486 [physics.ins-det].
\bibitem{oblath} 
  N.~S.~Oblath [ for the Project 8 Collaboration],
  arXiv:1310.0397 [physics.ins-det].
\bibitem{kunde} G.~J.~Kunde, DPF Presentation ID 296
\bibitem{chaves}  J. Chaves, DPF Presentation ID 99  and
  J.~B.~Albert {\it et al.}  [EXO-200 Collaboration],
  arXiv:1306.6106 [nucl-ex].
\bibitem{kravitz}  S. Kravitz, DPF presentation ID 83
\bibitem{giovanetti} G. Giovanetti, DPF presentation ID 250 and 
  N.~Abgrall {\it et al.}  [Majorana Collaboration],
  arXiv:1308.1633 [physics.ins-det].
\bibitem{szelc}  A. Szelc, DPF Presentation ID 95 and 
 O.~Palamara [ArgoNeuT Collaboration],
  J.\ Phys.\ Conf.\ Ser.\  {\bf 408}, 012039 (2013)
  [arXiv:1110.3070 [hep-ex]].
\bibitem{szelc2}  A. Szelc, DPF Presentation ID 164 and 
  P.~Adamson {\it et al.}  [LArIAT Collaboration],
  FERMILAB-PROPOSAL-1034.
\bibitem{grant} C. Grant, DPF Presentation ID 259 and 
  H.~Berns {\it et al.}  [ The CAPTAIN Collaboration],
  arXiv:1309.1740 [physics.ins-det].
\bibitem{laza}  L. Rakotondravohitra, DPF presentation ID 166 and 
  G.~A.~Fiorentini {\it et al.}  [MINERvA Collaboration],
  Phys.\ Rev.\ Lett.\  {\bf 111}, 022502 (2013)
  [arXiv:1305.2243 [hep-ex]].
and
  L.~Fields {\it et al.}  [MINERvA Collaboration],
  Phys.\ Rev.\ Lett.\  {\bf 111}, 022501 (2013)
  [arXiv:1305.2234 [hep-ex]].
\bibitem{bodek}  A. Bodek, DPF presentation ID 263 and 
  A.~Bodek, H.~S.~Budd and M.~E.~Christy,
  Eur.\ Phys.\ J.\ C {\bf 71}, 1726 (2011)
  [arXiv:1106.0340 [hep-ph]].
\bibitem{adam} J. Adam, DPF Presentation ID 229
\bibitem{hansen}  D. Hansen, DPF Presentation ID 238
\bibitem{worcester} E.~Worcester [Daya Bay Collaboration],
  arXiv:1309.7991 [hep-ex].
\bibitem{carr}  R. Carr, DPF Presentation ID 77 and 
  Y.~Abe {\it et al.}  [Double Chooz Collaboration],
  Phys.\ Lett.\ B {\bf 723}, 66 (2013)
  [arXiv:1301.2948 [hep-ex]].
\bibitem{hignight} J. Hignight, DPF Presentation ID 239 
\bibitem{radovic} A. Radovic, DPF presentation ID 0 and 
  P.~Adamson {\it et al.}  [MINOS Collaboration],
  Phys.\ Rev.\ Lett.\  {\bf 110}, 251801 (2013)
  [arXiv:1304.6335 [hep-ex]].
\bibitem{friend} M. Friend, DPF Presentation ID 268
\bibitem{sachdev}   K.~Sachdev,
  arXiv:1310.0119 [hep-ex].
\bibitem{baird} M. Baird, DPF Presentation ID 189 and 
  G.~Pawloski [NOvA Collaboration],
  Nucl.\ Phys.\ Proc.\ Suppl.\  {\bf 229-232}, 439 (2012).
\bibitem{bian}   J.~Bian,
  arXiv:1309.7898 [physics.ins-det].
\bibitem{collin} B. Carls, DPF Presentation ID 168 and 
  L.~Camilleri [MicroBooNE Collaboration],
  Nucl.\ Phys.\ Proc.\ Suppl.\  {\bf 237-238}, 181 (2013).
\bibitem{bass} M. Bass, DPF Presentation ID 256
\bibitem{guardincerri}  E. Guardincerri, DPF Presentation ID 258
\bibitem{shaevitz} M. Shaevitz, DPF Presentation ID 141 and 
  J.~M.~Conrad, M.~H.~Shaevitz, I.~Shimizu, J.~Spitz, M.~Toups and L.~Winslow,
  arXiv:1307.5081 [hep-ex].
\bibitem{spitz} 
  J.~S.~Díaz, T.~Katori, J.~Spitz and J.~M.~Conrad,
  arXiv:1307.5789 [hep-ex].
and 
  Y.~Abe {\it et al.}  [Double Chooz Collaboration],
  Phys.\ Rev.\ D {\bf 86}, 112009 (2012)
  [arXiv:1209.5810 [hep-ex]].
\bibitem{zhuridov} 
  D.~Zhuridov,
  arXiv:1309.2540 [hep-ph].
\bibitem{ahmed} R. Ahmed, DPF Presentation ID 220 
\bibitem{SNO} 
  S.~N.~Ahmed {\it et al.}  [SNO Collaboration],
  Phys.\ Rev.\ Lett.\  {\bf 92}, 181301 (2004)
  [nucl-ex/0309004].
\end{thebibliography}
\end{document}